\documentclass[preprint]{iucr}              

     \journalcode{A}              

\usepackage{graphicx,amssymb,amsmath,algpseudocode,algorithm,tabularx,verbatim,amsthm,lineno,color,physics}

\newtheorem{thm}{Theorem}
\newtheorem{dfn}[thm]{Definition}
\newtheorem{alg}[thm]{Algorithm}
\newtheorem{pro}[thm]{Problem}

\newtheorem{lem}[thm]{Lemma}
\newtheorem{exa}[thm]{Example}
\newtheorem{rmk}[thm]{Remark}

\newcommand{\R}{\mathbb R}
\newcommand{\Z}{\mathbb Z}
\newcommand{\al}{\alpha}
\newcommand{\be}{\beta}
\newcommand{\ga}{\gamma}

\newcommand{\de}{\delta}
\newcommand{\ep}{\epsilon}

\newcommand{\La}{\Lambda}
\newcommand{\id}{\mathrm{id}}
\newcommand{\Ker}{\mathrm{Ker}}
\newcommand{\Image}{\mathrm{Im}}

\newcommand{\vol}{\mathrm{vol}}
\newcommand{\MST}{\mathrm{MST}}
\newcommand{\SNF}{\mathrm{SNF}}

\newcommand{\LQG}{\mathrm{LQG}}
\newcommand{\vect}[2]{ \left( \begin{array}{c} 
 #1 \\ #2 \end{array} \right)}
\newcommand{\mat}[4]{ \left( \begin{array}{cc} 
 #1 & #2 \\ #3 & #4 \end{array} \right)}

\begin{document}                  



\title{Computing the bridge length: the key ingredient in a continuous isometry classification of periodic point sets}
\shorttitle{Computing the bridge length of a periodic point set}


\cauthor[a]{Jonathan}{McManus}{\{j.d.mcmanus,vkurlin\}@liverpool.ac.uk}{}
\author[b]{Vitaliy}{Kurlin}

\aff[a]{Computer Science department and Materials Innovation Factory, University of Liverpool, Liverpool L69 3BX \country{UK}}









\maketitle                        

\begin{synopsis}
We describe an efficient algorithm to compute the bridge length estimating the size of a complete isoset invariant, which classifies all periodic point sets under Euclidean motion.
\end{synopsis}

\begin{abstract}
The fundamental model of any periodic crystal is a periodic set of points at all atomic centres. 
Since crystal structures are determined in a rigid form, their strongest equivalence is rigid motion (composition of translations and rotations) or isometry (also including reflections).
The recent classification of periodic point sets under rigid motion used a complete invariant isoset whose size essentially depends on the bridge length, defined as the minimum `jump' that suffices to connect any points in the given set.
\smallskip

We propose a practical algorithm to compute the bridge length of any periodic point set given by a motif of points in a periodically translated unit cell. 
The algorithm has been tested on a large crystal dataset and is required for an efficient continuous classification of all periodic crystals. 
The exact computation of the bridge length is a key step to realising the inverse design of materials from new invariant values.

\keyword{mathematical crystallography}\keyword{periodic point set}\keyword{quotient graph}
\end{abstract}


\section{Introduction: practical motivations and the problem statement}
\label{sec:intro}

All solid crystalline materials can be modelled at the atomic level as periodic sets of points (with the chemical attributes if desired) at all atomic centres, defined below. 
 
\begin{figure}
\centering
\includegraphics[width=\textwidth]{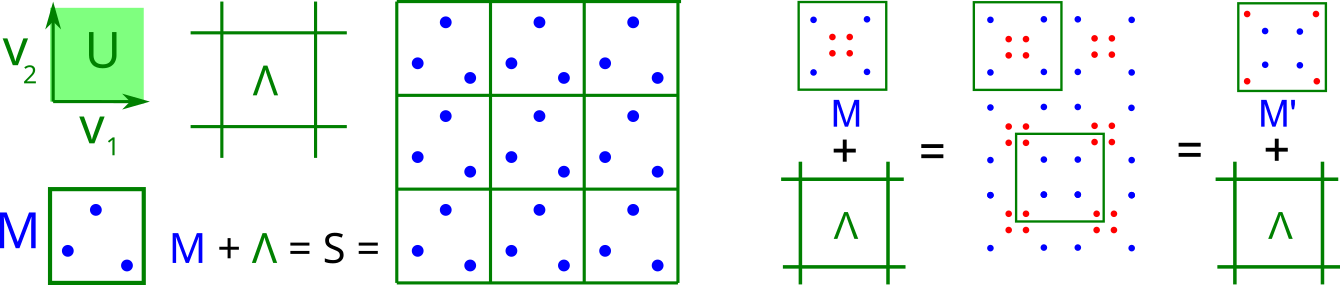}
\caption{\textbf{Left}: the orthonormal basis $\vb*{v_1},\vb*{v_2}$ generates
the green lattice $\La$ and the unit cell $U$ containing the blue motif $M$ of three points.
The periodic point set $S=\La+M$ is obtained by periodically repeating $M$ along all vectors of $\La$.
\textbf{Right}: different motifs $M,M'$ in the same cell generate periodic sets that differ by only translation.}
\label{fig:cell+motif=crystal}
\end{figure}

\begin{dfn}[lattice, unit cell, motif, periodic point set]
\label{dfn:periodic}
Any vectors $\vb*{v_1},\dots,\vb*{v_n}$ that form a linear basis of $\R^n$ generate
the \emph{lattice} $\La=\{\sum\limits_{i=1}^n c_i \vb*{v_i} \mid c_i\in\Z\}$ and the \emph{unit cell} $U=\{\sum\limits_{i=1}^n t_i \vb*{v_i} \mid 0\leq t_i<1\}$.
A \emph{motif} is any finite set of points $M\subset U$, which can represent centres of atoms in a real crystal.
The \emph{motif size} $|M|$ is the number of points in $M$. 
A \emph{periodic point set} $S=\La+M=\{\vb*{v}+p \mid \vb*{v}\in\La, p\in M\}$ is a union of $|M|$ lattices whose origins are shifted to all points $p$ of the motif $M$, see Fig.~\ref{fig:cell+motif=crystal}~(left). 
\end{dfn}

Any unit cell $U$ is a parallelepiped on basis vectors $\vb*{v_1},\dots,\vb*{v_n}$.
If we translate the unit cell $U$ by all vectors $v\in\La$, the resulting cells tile $\R^n$ without overlaps. 
Motif points represent atomic centres in a real crystal.
The same lattice can be generated by infinitely many different bases that are all related under multiplication by $n\times n$ matrices with integer elements and determinant 1. 
Even if we fix a basis of $\R^n$ and hence a unit cell $U$, different motifs in $U$ can define periodic point sets that differ only by Euclidean \emph{isometry} defined as any distance-preserving transformation of $\R^n$. 
\smallskip

Since crystal structures are determined in a rigid form, their slightly stronger equivalence is \emph{rigid motion} defined as any orientation-preserving isometry without reflections or as a composition of translations and rotations.
After many years of discussing definitions of a ``crystal'' \cite{brock2021change}, a \emph{crystal structure} was recently described in the periodic case as a class of periodic sets under rigid motion \cite{anosova2024importance}.
\smallskip

Any such class consists of all (infinitely many) periodic point sets that are equivalent to each other under some rigid motions.
However, almost any perturbation of atoms disturbs some inter-atomic distances and hence the isometry class with all cell-based descriptors, such as symmetry groups.
Even in dimension~1, for any integer $m>0$ and a small threshold $\ep>0$, the sequence $\Z$ with period 1 is pointwise $\ep$-close to the sequence with the motif $M=\{0,1+\ep,\dots,m+\ep\}$ and arbitrarily large period $m+1$. 
\smallskip

This inherent discontinuity of all cell-based descriptors was resolved by Pointwise Distance Distributions (PDD) in \cite{widdowson2022average,widdowson2022resolving,widdowson2026pointwise}, which defined geographic-style coordinates on the Cambridge Structural Database (CSD) in \cite{widdowson2024continuous}.
Though PDDs distinguish all periodic crystals in the CSD within minutes on a modest desktop, the only known theoretically complete and continuous invariant that uniquely identifies any periodic point set under isometry in $\R^n$ in polynomial time of the motif size (for a fixed dimension) is the \emph{isoset} \cite{anosova2021isometry, anosova2025 recognition}. 
\smallskip

The invariant isoset requires the bridge length whose definition is reminded below.

\begin{dfn}[bridge length $\be(S)$]
\label{dfn:bridge_length}
For any finite or periodic set of points $S\subset\R^n$, the \emph{bridge length} $\beta(S)$ is the minimum distance such that any points $p,q\in S$ can be connected by a finite sequence of points $p=p_1,p_2,\dots,p_k=q$ in $S$, such that every Euclidean distance has the upper bound $|p_i-p_{i+1}|\leq\beta(S)$ for all $i=1,\dots,k-1$.
\end{dfn}

Equivalently, the \emph{bridge length} $\be(S)$ is the minimum double radius such that the union of the closed balls of the radius $\frac{1}{2}\beta(S)$ around all points of $S$ is connected.
The lattice $\La=\Z^3$ of all points with integer coordinates has $\be(\La)=1$.
If we add to $\Z^3$ all points whose all coordinates are half-integer, the resulting BCC (body-centred cubic) periodic point set 
has $\be=\frac{\sqrt{3}}{2}$ equal to the half-diagonal of the unit cube in $\R^3$.
\smallskip

Expanding Delone's local theory \cite{delone1976local,dolbilin1976local,dolbilin1998multiregular,dolbilin2015delone,dolbilin2018delone}, Dolbilin and Bouniaev studied more general $t$-bonded Delone sets, where $t$ is an upper bound of the bridge length $\be(S)$ for any periodic point set $S\subset\R^n$, see \cite{bouniaev2017regular, dolbilin2019regular}.
The main problem below asks for an efficient algorithm to exactly compute $\be(S)$.

\begin{pro}
\label{pro:bridge_length}
Design an algorithm to compute the bridge length $\be(S)$ in polynomial time of the motif size for any periodic point set $S$ with a fixed unit cell in $\R^n$.
\end{pro}

\begin{figure}
\includegraphics[width=\textwidth]{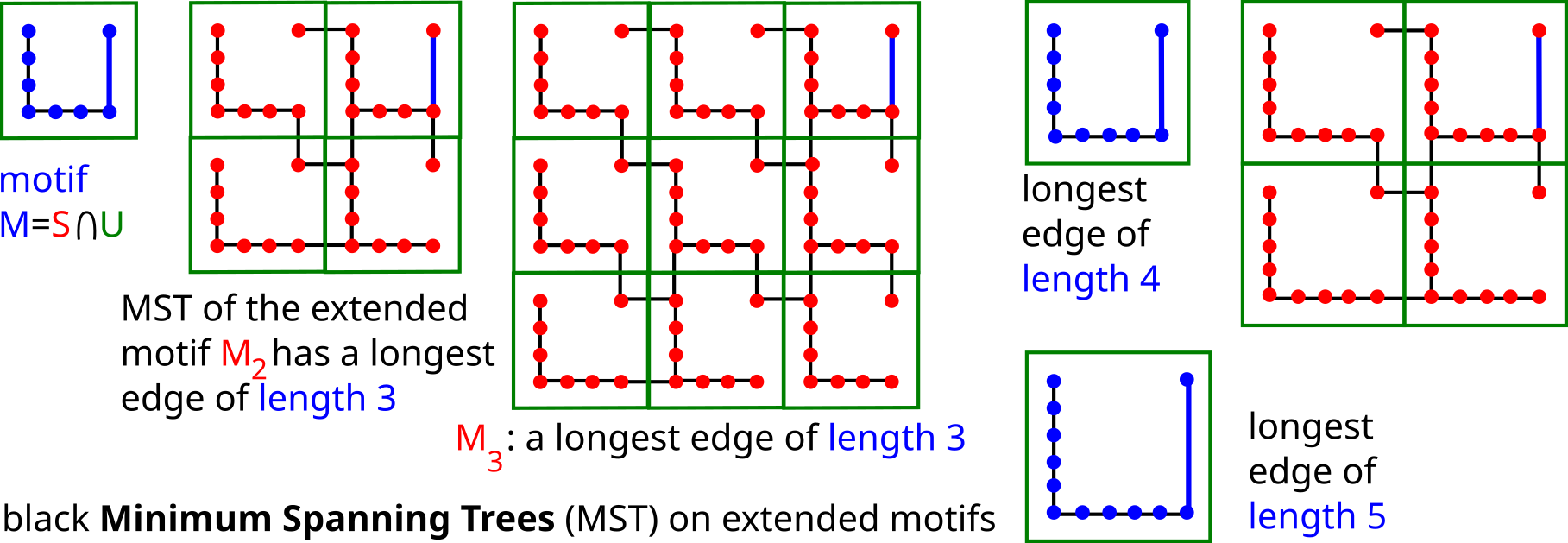}
\caption{All Minimum Spanning Trees on extended motifs of a periodic point set S have the longest edge (in blue) of length 3, which could be made arbitrarily long, relative to a preserved minimum inter-point distance of 1 and bridge length $\be(S)=2$ due to shorter edges from the top right point in every cell across a cell boundary.
}
\label{fig:bridge_length}
\end{figure}

The bridge length of a finite set can be computed via a Minimum Spanning Tree below, but the periodic case does not easily reduce to a finite one as shown in Fig.~\ref{fig:bridge_length}.

\begin{dfn}[Minimum Spanning Tree]
\label{dfn:MST}
For any finite set $M$ of points in $\R^n$, a \emph{Minimum Spanning Tree} $\MST(M)$ is a tree that has the vertex set $M$ and a minimum total length of straight-line edges with lengths measured by Euclidean distance.
\end{dfn}

$\MST(M)$ is uniquely defined if all distances between points of $M$ are distinct, see section 4.3 in \cite{sedgewick2011algorithms}.
By Definition~\ref{dfn:bridge_length}, the bridge length $\be(M)$ of any finite set $M\subset\R^n$ equals the length of the longest edge of $\MST(M)$.

For any periodic point set $S$ with a unit cell $U$ on a basis $\vb*{v_1},\dots,\vb*{v_n}$ in $\R^n$, one can consider the extended motifs $M_k=S\cap U_k$, where the extended cell $U_k$ is defined by the basis $k\vb*{v_1},\dots,k\vb*{v_n}$ for any integer $k>1$.
The Minimum Spanning Trees provide the upper bounds $\be(S)\leq\be(M_k)$ for $k>1$, which can be unnecessarily high, see Fig.~\ref{fig:bridge_length}, so Problem~\ref{pro:bridge_length} is much harder for periodic sets than for finite sets of points.
\smallskip

\begin{dfn}[parameters $r(U)$, $R(S)$, $a(U)$]
\label{dfn:parameters}
Let $S\subset\R^n$ be periodic point set whose a unit cell $U$ has a basis $\vb*{v_1},\dots,\vb*{v_n}$.
Set $r(U)=\max\{b,\frac{d}{2}\}$, where $b=\max\limits_{i=1,\dots,n}|\vb*{v_i}|$ and $d=\sqrt{v_1^2+\dots+v_n^2}$. 
The \emph{covering radius} $R(S)$ is the smallest radius $R$ such that the union of closed balls of radius $R$ around all $p\in S$ covers $\R^n$.
The \emph{height} is $h(U)=\vol(U)/\max\limits_{i=1,\dots,n}\vol(U_i)$, where $U_i$ is the subcell of $U$ spanned by all basis vectors except $\vb*{v_i}$.
The \emph{aspect ratio} of the cell $U$ is defined as $a(U)=r(U)/h(U)$. 
\end{dfn}

For any periodic set $S\subset\R^n$, Theorem~2 in \cite{delone1973three} and Lemma~3.6(a) in \cite{anosova2025recognition} imply the upper bound $\be(S)\leq \min\{r(U),2R(S)\}$, which is too high in practice, see section~\ref{sec:experiments}.
Main Theorem~\ref{thm:bridge_length_time} guarantees an exact computation of $\be(S)$ in a time that only quadratically depends on the motif size $m$ of $S$. 

\begin{thm}
\label{thm:bridge_length_time}
For any periodic point set $S \subset \mathbb{R}^n$ with a motif of $m$ points in a unit cell $U$, the bridge length $\be(S)$ can be computed in time $O(m^2 a(U)^nN)$, where $N$ is the time complexity of the Smith Normal Form, $a(U)$ is the aspect ratio from Definition~\ref{dfn:parameters}.
\end{thm}

As the time complexity is proportional to the aspect ratio $a(U)$ of a cell $U$, an initial reduction of $U$ to a smaller cell will speed up the computation of the bridge length by minimising further cell extensions, namely $supercell\_size$ in Algorithm \ref{alg:bridge_length}.
\smallskip

Section~\ref{sec:definitions} introduces the key concepts.
Section~\ref{sec:algorithm} describes the main algorithm for $\be(S)$.
Section~\ref{sec:time} proves Theorem~\ref{thm:bridge_length_time}.
Section~\ref{sec:experiments} presents experiments on crystals.

\section{Auxiliary concepts of graph theory for the bridge length algorithm}
\label{sec:definitions}

This section introduces a few auxiliary concepts to describe the exact algorithm for the bridge length $\be(S)$ in section~\ref{sec:algorithm} and to prove Theorem~\ref{thm:bridge_length_time} at the end of section~\ref{sec:time}.

\begin{figure}
\includegraphics[width=\textwidth]{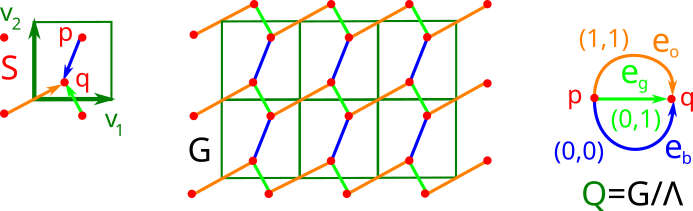}
\caption{\textbf{Left}: the periodic point set $S$ with the basis vectors $\vb*{v_1}=(5,0)$, $\vb*{v_2}=(0,5)$ and motif points $p=(2,1)$, $q=(3,4)$.
\textbf{Middle}: the periodic Euclidean graph $G\subset\R^2$ with three types of straight-line edges: green, blue, orange of lengths $\sqrt{5},\sqrt{10},\sqrt{20}$, respectively.
\textbf{Right}: the labelled quotient graph $Q$ has directed edges $e_g,e_b,e_o$ with translational vectors indicating integer shifts of cells, see Definitions~\ref{dfn:periodic_graph}, \ref{dfn:quotient_graph}, \ref{dfn:LQG}.}
\label{fig:quotient_periodic_graph}
\end{figure}

\begin{dfn}[$G\subset\R^n$] 
\label{dfn:periodic_graph}
Let $S\subset\R^n$ be a periodic point set with a lattice $\Lambda$.
A \emph{periodic Euclidean graph} $G\subset\R^n$ is an infinite graph with the vertex set $S$ and straight-line edges such that the translation by any vector $\vb*{v}\in\La$ defines an \emph{automorphism} of $G$, which is a bijection $S\to S$ that also induces a bijection on the edges of $G$, see Fig.~\ref{fig:quotient_periodic_graph}.
\end{dfn}

If straight-line edges 
meet at interior points, they are not considered vertices of $G$.

Fig.~\ref{fig:quotient_periodic_graph} shows a connected periodic graph $G$ but $G$ can also be disconnected.
For example, let $S$ be the square lattice $\Z^2$, then the graph $G$ consisting of all horizontal edges connecting the points $(m,n)$ and $(m+1,n)$ for $m,n\in\Z$ is periodic but not connected.
If we add to $G$ all vertical edges connecting $(m,n)$ and $(m,n+1)$ for $m,n\in\Z$, the resulting infinite square grid is a connected periodic graph on $\Z^2$.

\begin{dfn}[quotient graph] 
\label{dfn:quotient_graph}
Let $G$ be a periodic graph on a periodic point set $S$ with a lattice $\Lambda$ in $\mathbb{R}^n$.
Two points of $S$ (also vertices or edges of $G$) are called \textit{$\Lambda$-equivalent} if they are related by a translation along a vector $\vb*{v}\in\La$. 
The \emph{quotient graph} $G/\La$ is an abstract undirected graph obtained as the quotient of $G$ under the $\Lambda$-equivalence.
Then $G$ is called a \emph{lifted graph} of $G/\La$.
Any vertex of $G/\La$ is a $\La$-equivalence class $p+\La$ represented by a point $p\in S$. 
Any edge $e$ of the quotient graph $G/\La$ is a $\La$-equivalence class $[p,q]+\La$ represented by a straight-line edge $[p,q]$ of $G$.
\end{dfn}

The quotient graph $G/\La$ can have multiple edges between the same pair of vertices as shown in Fig.~\ref{fig:quotient_periodic_graph}, which all can be distinguished by the labels defined below.

\begin{dfn}[labelled quotient graph]
\label{dfn:LQG}
Let $S\subset\R^n$ be a periodic point set with a lattice $\La$ defined by a basis $\vb*{v_1},\dots,\vb*{v_n}$.
Let $G$ be a periodic graph on $S$.
For an edge $e$ of the quotient graph $G/\La$, choose any of two directions and a representative edge $[p,q]$ in the lifted graph $G$.
Let $U(p),U(q)$ be the unit cells containing $p,q$, respectively.
There is a unique vector $\vb*{v}=\sum\limits_{i=1}^n c_i\vb*{v_i}\in\La$ such that $U(q)=U(p)+v$ and $c_i\in\Z$, and we define the length of edge $e$ in $G/\La$ as the Euclidean distance $|p-q|$. 
This 'length' of an edge $e$ is considered an attribute to ease later calculations, and does not change the abstract nature of the quotient graph $G/\La$.
\smallskip

A \emph{labelled quotient graph ($\LQG$)} is $G/\La$ whose every edge $e$ has a direction (say, from the $\La$-equivalence class of $p$ to $\La$-equivalence class of $q$) and the \emph{translational vector} $\vb*{v}(e)=(c_1,\dots,c_n)\in\Z^n$, see Fig.~\ref{fig:quotient_periodic_graph}. 
Changing the direction of $e$ multiplies each coordinate of $\vb*{v}(e)$ by $(-1)$.
An equivalence of LQGs is a composition of a graph isomorphism and changes in edge directions that match all translational vectors.
\end{dfn}

Translational vectors $\vb*{v}(e)$ are also called \emph{voltages} if $G/\La$ is considered a \emph{voltage graph} or a \emph{gain graph} in topological graph theory.
In crystallography, labelled quotient graphs have been studied by many authors.
Section~6 in \cite{chung1984nomenclature} generated 3-periodic nets by considering LQGs whose translational vectors have entries from $\{-1,0,1\}$. 
Section~2 in \cite{cohen1990recognizing} described an algorithm to find connected components of a fixed periodic graph in terms of its LQG.
Proposition~5.1 in \cite{eon2011euclidean} showed how to reconstruct a periodic graph up to translations from LQG and a lattice basis, which we also prove in Lemma~\ref{lem:lifting} in our notations for completeness.
Section~3 in \cite{eon2016topological} described surgeries on building units of LQGs.
Theorem~6.1 in \cite{eon2016vertex} characterised 3-connected minimal periodic graphs (with a slightly different definition of 'minimal').
\cite{mccolm2024realizations} initiated a search for systematic periodic graphs realisable by real crystal nets, see also
\cite{edelsbrunner2024merge}.
\smallskip

The labelled quotient graph $G/\La$ in Fig.~\ref{fig:quotient_periodic_graph} has two vertices $p,q$.
If we orient the three edges of $Q=G/\La$ from $p$ to $q$, the translational vector $(0,0)$ of the blue edge $e_b$ in $G/\La$ means that the corresponding straight-line blue edge in the lifted graph $G\subset\R^2$ connects points of $S$ within the same unit cell $U$ with the basis $\vb*{v_1},\vb*{v_2}$.
The orange edge with the translational vector $(1,1)$ means that each of its infinitely many liftings in $G\subset\R^2$ joins a point in a cell $U$ to another point in the cell $U+\vb*{v_1}+\vb*{v_2}$. 

\begin{lem}[lifting]
\label{lem:lifting}
Let $G$ be a periodic Euclidean graph on a periodic point set $S$ with a motif $M$ in a unit cell $U$ defined by a basis $\vb*{v_1},\dots,\vb*{v_n}$ in $\R^n$.
Let $Q$ be a labelled quotient graph of $G$. 
Then $G\subset\R^n$ can be reconstructed from $Q$, the basis $\vb*{v_1},\dots,\vb*{v_n}$, and a bijection between all vertices of $Q$ and all points of the motif $M\subset U$.
\end{lem}
\begin{proof}
The basis $\vb*{v_1},\dots,\vb*{v_n}$ is needed to define a unit cell $U$ with the given points of $M$, which are in 1-1 correspondence with all vertices of $Q$.
The periodic point set $S$, which is the vertex set of the periodic graph $G$, is obtained from $M$ by translations along the vectors $\sum\limits_{i=1}^n c_i \vb*{v_i}$ for all $c_i\in\Z$.
By Definitions~\ref{dfn:quotient_graph} and~\ref{dfn:LQG}, every edge $e$ of the labelled quotient graph $Q$ has a translational vector $\vb*{v}(e)=(c_1,\dots,c_n)$ and is a $\La$-equivalence class $[p,q]+\La$ for some $p,q\in S$ whose unit cells $U(p),U(q)$ are related by the translation along $\sum\limits_{i=1}^n c_i \vb*{v_i}$.
Then we can lift the edge $e$ to the periodically translated straight-line edges $[p+\vb*{v},q+\vb*{v}+\sum\limits_{i=1}^n c_i \vb*{v_i}]$ in the periodic graph $G$ for all $\vb*{v}\in\La$.
\end{proof}


\begin{dfn}[path/cycle sum]
\label{dfn:cycle_sum}
For a path (sequence of consecutive edges) in a labelled quotient graph $Q$, we make all directions of edges consistent in the sequence and define the \emph{path sum} in $\Z^n$ as the sum of the resulting translational vectors along the path. 
If the path is a closed cycle, the path sum is called the \emph{cycle sum}.
\end{dfn}

In the language of \emph{voltage graphs}, a path sum 
may equivalently be referred to as the \emph{net voltage} over the path.
In the labelled quotient graph in Fig.~\ref{fig:quotient_periodic_graph}, the upper cycle consisting of the directed orange edge (from $p$ to $q$) and the inverted green edge (from $q$ to $p$) has the cycle sum $(1,1)+(0,-1)=(1,0)$.
This cycle sum means that a lifting of the cycle to the periodic graph $G$ in $\R^2$ produces a polygonal path connecting a point to its translate by the vector $\vb*{v_1}=(1,0)$ in the next cell to the right.

\begin{dfn}[minimal tree $\MST(S/\La)$]
\label{dfn:tree}
For a periodic point set $S\subset\R^n$ with a lattice $\La$, a \emph{minimal tree} is a Minimum Spanning Tree $\MST(S/\La)$ (Definition~\ref{dfn:MST}) on the set $S/\La$ of $\La$-equivalence classes of points, where the distance between any classes in $S/\La$ is the minimum Euclidean distance between their representatives in the set $S$. 
\end{dfn}

In Fig.~\ref{fig:quotient_periodic_graph}, a minimal tree $\MST(S/\La)$ consists of one shortest green edge in $G/\La$.

\section{Algorithm for the bridge length of a periodic point set}
\label{sec:algorithm}

This section will describe main Algorithm~\ref{alg:bridge_length} for solving Problem~\ref{pro:bridge_length}, which will call auxiliary Algorithm~\ref{alg:edges} several times.
Algorithm~\ref{alg:edges} starts from a conventional representation of a periodic set $S\subset\R^n$ with a motif $M$ of points given by coordinates in a basis $\vb*{v_1},\dots,\vb*{v_n}$ of a lattice $\La$ as in a Crystallographic Information File (CIF).
\smallskip

At every call, Algorithm~\ref{alg:edges} returns the next shortest edge $e$ between points of $S$ in increasing order of length.
Although $S$ is a set of points rather than a graph, we will use the term 'edge', because $e$ can be considered an edge from a complete graph with the vertex set $S$ and with the 'next shortest edge' being up to $\Lambda$-equivalence. 
\smallskip

Any edge $e$ between points of $S$ will be represented by an ordered pair of points $p,q\in M$ and a translational vector $(c_1,\dots,c_n)\in\Z^n$ so that the actual straight-line edge in the lifted periodic graph $G\subset\R^n$ is from $p$ to the point $q+\sum\limits_{i=1}^n c_i \vb*{v_i}$.
For convenience, we record the Euclidean distance $d=|q-p+\sum\limits_{i=1}^n c_i \vb*{v_i}|$ between these endpoints.
Then Algorithm~\ref{alg:edges} outputs any edge $e$ as a tuple $(p,q;c_1,\dots,c_n;d)$.
\smallskip

Algorithm~\ref{alg:edges} maintains the list of already found edges in increasing order of length.
If the next required edge $e$ is already in the list, Algorithm~\ref{alg:edges} simply returns $e$.
This shortcut is implemented in Python with the keyword `Yield', see the documentation at 
https://docs.python.org/3/glossary.html\#term-generator-iterator.
Rather than starting from line 1, every time when Algorithm~\ref{alg:edges} is called, each call 'Yield $e$' returns an edge $e$, then temporarily suspends processing, remembering the location execution state including all local variables. 
When 'Yield $e$' is called again, Algorithm~\ref{alg:edges} picks up where it left off in contrast to functions that start fresh on every invocation.
\smallskip

If the next edge $e$ is not yet found, Algorithm~\ref{alg:edges} adds more points from a shell of unit cells surrounding the previously considered cells.
This \emph{shell} contains the extended motif $M_k$ without the smaller motif $M_{k-1}$ for $k>1$, see Fig.~\ref{fig:bridge_length}.
For any new point $p$, it suffices to consider only edges to points $q\in M\subset U$ because any edge $e$ can be periodically translated by $\vb*{v}\in\La$ so that one of the endpoints of $e$ belongs to $U$.
In Algorithm~\ref{alg:edges}, the \emph{Chebyshev distance} $\ell_\infty$ in line 3 is the maximum absolute difference of corresponding coordinates, while $d$ in line 7 is the usual \emph{Euclidean distance}.
\smallskip

\begin{alg}
    \label{alg:edges}
    Input: a basis $\vb*{v_1},\dots,\vb*{v_n}$ defining a unit cell $U$, a motif $M\subset U$. \\
    \textbf{\emph{next\_edge} runs only until the next \emph{Yield}, and outputs the yielded edge}.
    \begin{algorithmic}[1]
        \State supercell\_size=0,
        current\_batch=[],
        next\_batch=[],
        next\_batch\_min\_length=$\infty$
        \While{True}
        \For{transl\_vector\ in $\mathbb{Z}^n$ s.t. $\ell_\infty(\vb*{0},transl\_vector)=supercell\_size$}
                \For{source in the motif M}
                    \For{dest in the motif M}
                        \State $true\_dest=dest+basis \cdot transl\_vector$
                        \State $length = d(source, true\_dest)$
                        \State next\_batch.append((length, source, dest, transl\_vector))
                        \State next\_batch\_min\_length = min\{length,next\_batch\_min\_length\}
                    \EndFor
                \EndFor
            \EndFor
            \While{current\_batch is not empty}
                \State next\_length = minimum\_length(current\_batch)
                \If{next\_length $\geq$ next\_batch\_min\_length} Break
                \EndIf
                \State current\_batch.remove(next\_length)
                \State Yield(next\_length)
            \EndWhile
            \State current\_batch = concatenate(current\_batch, next\_batch)
            \State next\_batch=[]
            \State supercell\_size = supercell\_size + 1
        \EndWhile 
    \end{algorithmic}
\end{alg}

There is a faster way of checking a condition equivalent to $next\_batch\_min\_len$ by using the cell geometry. 
Then in the vast majority of cases the algorithm can stop at a supercell one size smaller, which dramatically speeds up the calculation. 
This calculation is described in Remark~\ref{rmk:alternative}. 
However, due to the possibility of that not being the case (upon which the algorithm would just default to the same supercell size), we will keep this simpler idea and use it for the time complexity calculations.

\begin{rmk}[a faster way to compute $next\_batch\_min\_len$ in Algorithm \ref{alg:edges}]
\label{rmk:alternative}
For a unit cell with a basis $\vb*{v_1},\dots,\vb*{v_n}$, 
let $\vb*a_{i}$ and $\vb*b_{i}$ be the shortest vectors parallel and antiparallel to $\vb*{v_i}$ from any point of a motif $M\subset U$ to the opposite boundary faces of the unit cell $U$.
Then the faster alternative for $next\_batch\_min\_len$ is 
\[
\min\limits_{i=1,\dots,n}(\vert\vb*{a_i}\vert+\vert\vb*{b_i}\vert+supercell\_size*\vert\vb*{v_i}\vert).
\]
As all the vector lengths $\vert \vb*{a_i}\vert,\vert\vb*{b_i}\vert$, $i=1,\dots,n$ can be pre-computed, we get a massive improvement over the calculation of $next\_batch\_min\_len$ in Algorithm \ref{alg:edges}.
\end{rmk}
\smallskip

Algorithm~\ref{alg:bridge_length} will be building a labelled quotient graph $Q$ by adding (or ignoring) edges found by Algorithm~\ref{alg:edges} and monitoring the connectivity of the growing lifted graph $G$ whose quotient $G/\La$ is $Q$.
For a basis $\vb*{v_1},\dots,\vb*{v_n}$ of a unit cell $U$ of the lattice $\La$ of $S$, the edge $e$ between points $p$ and $q+\sum\limits_{i=1}^n c_i\vb*{v_i}\in S$ is added to $Q$ as the edge between the $\La$-equivalence classes of $p$ and $q$, with the translational vector $\vb*{v}(e)=(c_1,\dots,c_n)\in\Z^n$. 
As soon as $G$ becomes connected, the length of the last added edge is the bridge length $\be(S)$, which will be proved in Theorem~\ref{thm:bridge_length_algorithm} later.
\smallskip

\begin{figure}[]{}
  \centering
  \includegraphics[width=\textwidth]{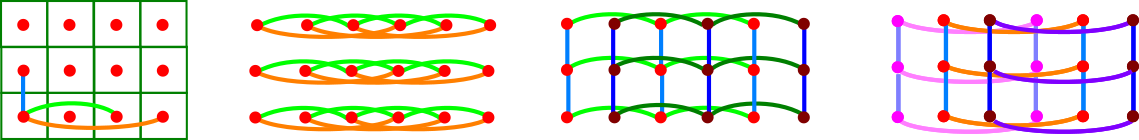}
  \caption{\textbf{Left}: the 3 vectors $\vb*{v_1}=(0,1)$, $\vb*{v_2}=(2,0)$, $\vb*{v_3}=(3,0)$ form a basis of $\Z^2$.
  \textbf{Other images}: none of the 3 pairs $(\vb*{v_2}, \vb*{v_3})$, $(\vb*{v_1}, \vb*{v_2})$, $(\vb*{v_1}, \vb*{v_3})$ form a basis (insufficient for full connectedness) of $\Z^2$.
Some straight edges are shown curved for better visibility.}
  \label{fig:Z2basis3vectors}
\end{figure}

In comparison with a Minimum Spanning Tree of a finite set of points, verifying the connectivity of the lifted periodic graph requires a much more complicated check that translational vectors with integer coordinates form a basis in $\Z^n$ (not $\R^n$), which can include more than $n$ vectors.
Fig. \ref{fig:Z2basis3vectors} shows a basis of $\Z^2$ consisting 3 vectors, where no vector can be dropped without losing the connectivity of all integer points in $\Z^2$. 
\smallskip

Algorithm~\ref{alg:bridge_length} will use the Smith Normal Form ($\SNF$) of a matrix of vectors $(c_1,\dots,c_n)$ in $\Z^n$, see p.~26 in \cite{newman1972integral}, \cite{cohn1985free}, and chapter~3.6 in \cite{van2003algebra} for finitely generated modules over a Principal Ideal Domain (PID).
 


\begin{dfn}[Smith Normal Form and invariant factors]
\label{dfn:SNF}
For integers $m,n\geq 1$, let $A$ be a non-zero $n\times m$ matrix over a Principal Ideal Domain $P$, for example, $P=\Z$. 
Then there exist invertible $n\times n$ and $m\times m$-matrices $L$,$R$, respectively, with coefficients in $P$, such that the product $LAR$ is an $n\times m$ matrix whose only non-zero entries are diagonal elements $a_i$ such that $a_i$ divides $a_{i+1}$ for $i=1,\dots,j-1$, and $a_i=0$ for $i=j,\dots,n$ for some $1<j\leq n$. 
This diagonal matrix $LAR$ is the \emph{Smith Normal Form} $\SNF(A)$.
The diagonal elements $a_i$ are called the \emph{invariant factors} of $A$.
\end{dfn}

Let $1$ denote the unit element of a Principal Ideal Domain $P$. 
If $P=\Z$, then $1$ is the usual integer 1.
The simplest SNF has all invariant factors equal to 1, which happens if and only if the last factor $a_n=1$ because all previous factors $a_i$ divide $a_n$.

\begin{alg}[Finding the bridge length $\be(S)$ of any periodic point set $S\subset\R^n$]
\label{alg:bridge_length}
\textbf{Initialisation}. 
A labelled quotient graph $Q$ and a forest $F\subset Q$ initially consist of $m$ isolated vertices, each representing a $\Lambda$-equivalence class of a point of the motif of $S$.
We will build a \emph{translational matrix} $A$ with columns in $\Z^n$, which is initially empty.
\medskip

\noindent
\textbf{Loop stage}. 
Consider the next edge $e=next\_edge()$ found by Algorithm~\ref{alg:edges}.
\smallskip

\noindent
\textbf{Case 1}.
If adding the edge $e$ to the current forest $F$ would not form a closed cycle (ignoring all edge directions), then add $e$ to $F$ and $Q$ as an edge with an arbitrarily chosen direction and corresponding translational vector $\vb*{v}(e)$ found by Algorithm~\ref{alg:edges}.
\smallskip

\noindent
\textbf{Case 2}.
If adding the edge $e$ to $F$ does form a cycle, find its cycle sum $c\in\Z^n$ from Definition~\ref{dfn:cycle_sum}. 
If $c$ is not $0\in\Z^n$ and cannot be expressed as an integer linear combination of the columns from the current translational matrix $A$, then add $e$ to $Q$ as in Case~1 (but not to the forest $F$) and add the vector $c$ as a new column to $A$.
\medskip
    
\noindent
\textbf{Termination}. 
Stop if both conditions below hold, otherwise continue the loop.

\noindent
(1) the labelled quotient graph $Q$ (hence the forest $F$) becomes connected; and
 
\noindent
(2) the translational matrix $A$ (whose columns are cycle sums of cycles created by adding edges) has $n$ invariant factors equal to 1, see Definition~\ref{dfn:SNF}.
\end{alg}
\smallskip

The necessity of termination condition~1 in Algorithm~\ref{alg:bridge_length} means that if the lifted periodic graph $G$ is connected then so is its quotient $Q=G/\La$.
The inverse implication (sufficiency) may not hold.
For example, in Fig.~\ref{fig:quotient_periodic_graph}, the minimal tree $\MST(S/\La)$ is a single green edge $e_g$, whose preimage under the quotient map $G\to G/\La$ is the disconnected set of all green straight-line edges in the periodic graph $G\subset\R^2$.
\smallskip

\begin{exa}[running Algorithm~\ref{alg:bridge_length} on the periodic point set $S$ in Fig.~\ref{fig:quotient_periodic_graph}]
\label{exa:bridge_length}
The first addition to the quotient graph $Q$ and forest $F$, which initially had two isolated vertices $p,q$, is the shortest green edge $e_g$ from $p$ to $q$ (case 1 in the loop stage) with the translational vector $c(e_g)=(0,1)\in\Z^2$.
The translational matrix $A$ remains empty.
\smallskip

Adding the next (by length) blue edge $e_b$ with $c(e_b)=(0,0)$ to $F=\{e_g\}$ creates a cycle with the cycle sum $c=c(e_g)-c(e_b)=(0,1)$.
According to case 2 in the loop stage, the quotient graph $Q$ becomes the cycle of two edges $e_g\cup e_b$ but the forest remains $F=\{e_g\}$.
The translational matrix $A$ becomes one column $\vect{0}{1}$ and does not yet have two invariant factors 1.
The 2nd termination condition is not yet satisfied, and the current lifted graph consisting of all green and blue segments is still disconnected. 
\smallskip

Adding the orange edge $e_o$ with $c(e_o)=(1,1)$ to $F$ creates another cycle with the cycle sum $c'=c(e_g)-c(e_o)=(-1,0)$.
The quotient graph $Q=e_g\cup e_b\cup e_o$ is now full but $F=\{e_g\}$ is still one edge.
The matrix $A$ becomes $\mat{0}{-1}{1}{0}$ whose $\SNF=\mat{1}{0}{0}{1}$ shows that $A$ has 2 invariant factors equal to 1.
Both termination conditions hold and the lifted periodic graph $G\subset\R^2$ of all green, blue, and orange edges is connected.
The bridge length $\be(S)=2\sqrt{5}$ equals the length of the last (orange) edge as expected.
\end{exa}

\section{Correctness and time complexity of the bridge length algorithm}
\label{sec:time}

This section proves the correctness of Algorithm~\ref{alg:bridge_length} in Theorem~\ref{thm:bridge_length_algorithm} about the bridge length and main Theorem~\ref{thm:bridge_length_time} about its time complexity.
Lemmas~\ref{lem:basis=factors1}-\ref{lem:factors1} will prove the necessity of termination condition~2 in Algorithm~\ref{alg:bridge_length}.
Both conditions 1 and 2 will guarantee the connectedness of the lifted periodic graph $G$ due to Lemma~\ref{lem:sufficiency}.


%
%

Lemma~\ref{lem:splitting} is a partial case of 
the splitting lemma on page 147 in \cite{hatcher2011AT}.

\begin{lem}[splitting]
\label{lem:splitting}
A short sequence of linear maps 
$0\to \Z^{m-n} \xrightarrow{f} \Z^{m} \xrightarrow{g} \Z^n \to 0$ is called \emph{exact} if the image of each map coincides with the \emph{kernel} (subspace mapping to 0) of the next map, i.e. $\Ker(f)=0$, $\Image(f)=\Ker(g)$, $\Image(g)=\Z^n$.
If there is a map $h: \Z^n \to \Z^{m}$, such that $g \circ h$ is the identity on $\Z^n$, then 
$\Z^{m}\cong f(\Z^{m-n})\oplus h(\Z^n)$, where $f(\Z^{m-n})$ and $h(\Z^n)$ are linearly independent subspaces of $\Z^m$ for $m\geq n$. 
\end{lem}

Note that if $f$ is a linear map, and $Ker(f)=\{\vb{0}\}$, then $f$ is injective, because  $f(x)=f(y)$ implies that $0=f(x)-f(y)=f(x-y)$, so $x-y=0$ and $x=y$.

\begin{exa}[finding a Smith Normal Form]
\label{exa:SNF}
In the notations of Lemma~\ref{lem:splitting}, Fig. \ref{fig:Z2basis3vectors} defines $g:\Z^3\to\Z^2$ given by the matrix 
$A=\begin{pmatrix}
1&0&0 \\
0&2&3
\end{pmatrix}$
whose 3 columns generate $\Z^2$.
The kernel $\Ker(g)\subset\Z^3$ consists of all vectors 
$f(k)=k\begin{bmatrix}
    0\\3\\-2
\end{bmatrix}$ for $k\in\Z$, which defines $f:\Z\to\Z^3$ with $\Ker(f)=0$ and $\Image(f)=\Ker(g)$ as required in Lemma~\ref{lem:splitting}.
Since $g:\Z^3\to\Z^2$ is surjective, we can find a map $h:\Z^2\to\Z^3$ satisfying $g\circ h=\id$, e.g. $h$ can be given by $M=
\begin{pmatrix}
    1&0\\0&-1\\0&1
\end{pmatrix}$, then
$A M = \begin{pmatrix}
    1&0&0\\
    0&2&3
\end{pmatrix}
\begin{pmatrix}
    1&0\\
    0&-1\\
    0&1
\end{pmatrix}=
\begin{pmatrix}
    1&0\\
    0&1
\end{pmatrix}$, denoted by $I_2$. 
After extending the $3\times 2$ matrix $M$ by the extra column with a basis vector of $\Image(f)$, we get the matrix $R=\begin{pmatrix}
    1&0&0\\
    0&-1&3\\
    0&1&-2
\end{pmatrix}$
such that $AR=\begin{pmatrix}
    1&0&0\\
    0&1&0
  \end{pmatrix}$. 
  
Lemma~\ref{lem:splitting} implies that the constituent blocks of $R$ are linearly independent to each other; all columns of $R$ are linearly independent, and $R$ is invertible. 
Hence, $I_2 A R$ is a Smith Normal Form of $A$ with $n=2$ invariant factors equal to 1 by Definition~\ref{dfn:SNF}.
\end{exa}

\begin{lem}[matrix generating $\Z^n$ $\Leftrightarrow$ $n$ invariant factors equal to 1]
\label{lem:basis=factors1}
The columns of any $n\times m$ matrix $A$ generate $\Z^n$ if and only if $A$ has $n$ invariant factors equal to 1.
\end{lem}
\begin{proof}
Let the $m$ columns of $A$ generate $\Z^n$. 
Then $A$ defines the surjection $g:\Z^{m}\to\Z^n$ whose kernel $\Ker(g)$ can be obtained as the image of a map $f:\Z^{m-n}\to\Z^m$, chosen such that $\Ker(g)$ is generated by $f(\vb*{e_1}),\dots,f(\vb*{e_{m-n}})$, where $\vb*{e_1},\dots,\vb*{e_{m-n}}$ denote the standard orthonormal basis of $\Z^{m-n}$.
Since $g:\Z^{m}\to\Z^n$ is surjective, orthonormal basis vectors $\vb*{u_1},\dots,\vb*{u_n}$ of $\Z^n$ are images $g(\vb*{v_1}),\dots,g(\vb*{v_n})$, respectively, of some vectors $\vb*{v_1},\dots,\vb*{v_n}\in\Z^m$.
We can define the linear map $h:\Z^{n}\to\Z^m$, $h(\vb*{u_i})=\vb*{v_i}$ for $i=1,\dots,n$, so that $g\circ h=\id$ on $\Z^n$.
Then $h$ has the $m\times n$ matrix $M$ such that $AM=I_n$, where $I_n$ is the $n\times n$ identity matrix.  
Extending $M$ by the $m-n$ columns $f(\vb*{e_1}),\dots,f(\vb*{e_{m-n}})$ gives the invertible $m\times m$ matrix $R$ such that $AR$ equals the $n\times m$ matrix obtained by extending $I_n$ with $m-n$ zero columns. 
Again, $R$ is an invertible matrix over $\Z$, so $I_m A R = A R$ is the Smith Normal Form of $A$ with all invariant factors equal to $1$ by Definition~\ref{dfn:SNF}.
\smallskip

Conversely, let the Smith Normal Form $\SNF=LAR$ of the matrix $A$ in Definition~\ref{dfn:SNF} have all invariant factors equal to 1, so the $n\times m$ matrix $LAR$ has the first $n$ columns $\vb*{u_1},\dots,\vb*{u_n}$, which form a standard basis of $\Z^n$, and $m-n$ zero columns.
Then each $\vb*{u_i}$ is a linear combination of the $m$ columns of $AR$ with coefficients from $L$, so these $m$ columns generate $\Z^n$.
Since 
$R$ is invertible, the $m$ columns of $A$ also generate $\Z^n$.    
\end{proof}

\begin{lem}[connected periodic graph $G\subset\R^n$ $\Rightarrow$ $n$ invariant factors equal $1$]
\label{lem:factors1}
In Algorithm~\ref{alg:bridge_length}, if the lifted periodic graph $G\subset\R^n$ whose vertices form a periodic set $S$ becomes connected, then the translational matrix $A$ has $n$ invariant factors equal to 1.
\end{lem}
\begin{proof}
By Lemma~\ref{lem:basis=factors1} it suffices to show that any vector $\vb*{v}\in\Z^n$ is an integer linear combination of columns of $A$.
Let $\La$ be the lattice of $S$ in Algorithm~\ref{alg:bridge_length} and $p\in S$ be any point.
The points $p$ and $p+\vb*{v}$ are connected in the lifted graph $G\subset\R^n$ by a polygonal path of straight-line edges.
Under $G\to G/\La$, this path projects to a closed cycle $C$ at the vertex ($\La$-equivalence class) $p+\La$ in the quotient graph $Q=G/\La$.
\smallskip

Let the cycle $C$ pass through edges $e_1,\dots,e_k$ (with integer multiplicities) in the complement $Q-F$ of the forest $F$ in the quotient graph $Q$. 
These edges were added only to $Q$ in case 2 of the loop stage.
When we tried to add every edge $e_j$ to $F$, the edge $e_j$ created a cycle $C_j$ whose cycle sum appeared as a column in the translational matrix $A$ (if this cycle sum was not yet an integer combination of the previous columns).
Then the vector $\vb*{v}$ equals the sum of the cycle sums of all the cycles $C_j$ for $j=1,\dots,k$, which is an integer combination of the columns of $A$ as required. 
\end{proof}

\begin{lem}[connected quotient graph $G/\La$ $\Rightarrow\exists$ a tree of representatives $T\subset G$]
\label{lem:tree_representatives}
If a labelled quotient graph $Q=G/\La$ is connected, its lifted graph $G\subset\R^n$ on a periodic point set $S$ with a motif of $m$ points and a lattice $\La$ includes a straight-line \emph{tree of representatives} $T\subset G$ with $m$ vertices that are not $\La$-equivalent to each other.
\end{lem}
\begin{proof}
Since $Q$ is connected, we can choose a spanning tree $F\subset Q$ on the $m$ vertices of $Q$.
A required tree $T\subset G$ will be a connected union of straight-line edges of $G$ that map 1-1 to all edges of $F$ under the quotient $G\to Q$.
Start from any point $p\in S$ and take any edge $e$ at the vertex ($\La$-equivalence class) $p+\La$ of $F\subset Q$.
The preimage of $e$ under $G\to Q$ contains a unique straight-line edge $[p,q]\subset G$, which we add to $T$.
After adding to $T$ all edges at $p$ that project to all edges of $F$ at the vertex $p+\La$,
choose another point $p'\in T$ such that the vertex $p'+\La$ has an edge of $F$ not yet covered by $T$ under $G\to Q$.
We continue adding edges to $T$ by using their projections in $F\subset Q$ until we get a tree $T\subset G$ that spans $m$ points of $S$ that are not $\La$-equivalent.
The final $T$ has no cycle, else this cycle projects under $G\to Q$ to a cycle in a forest $F$.
\end{proof}

\begin{lem}[termination conditions in Algorithm~\ref{alg:bridge_length} $\Rightarrow$ connected graph $G\subset\R^n$]
\label{lem:sufficiency}
Let $Q$ be a labelled quotient graph with a translational matrix $A$ and a lifted graph $G$ on a periodic point set $S\subset\R^n$ with a lattice $\La$.
If $Q$ is connected and the matrix $A$ has $n$ invariant factors equal to 1, then the lifted periodic graph $G\subset\R^n$ is connected.
\end{lem}
\begin{proof}
For any points $p,q\in S$, we will find a path of straight-line edges in $G$ as follows.
By Lemma~\ref{lem:tree_representatives} the connectedness of the quotient graph $Q=G/\La$ guarantees the existence of a tree $T\subset G$ whose vertices represent all $\La$-equivalences classes of points of $S$.
Let $p',q'$ be the vertices of $T$ that are $\La$-equivalent to $p,q$, respectively.
\smallskip

Since $p',q'$ are connected by a path in $T$, it suffices to find a path from $p$ to its $\La$-translate $p'=p+\vb*{v}$ (then similarly from $q$ to $q'$) in the graph $G$ for any $\vb*{v}\in\La$.
By Lemma~\ref{lem:basis=factors1} the columns of $A$ form a basis of $\Z^n$, so $\vb*{v}$ is an integer combination of these columns.
It suffices to find a path in $G$ by assuming that $\vb*{v}$ is one column of $A$ because a path for any sum $\sum_i\vb*{v}_i$ can be obtained by concatenating paths for $\vb*{v}_i$.
A column $\vb*{v}$ can appear in $A$ only in case 2 of the loop stage in Algorithm~\ref{alg:bridge_length} as a cycle sum of a cycle $C\subset Q$ that was created by trying to add an edge $e$ from Algorithm~\ref{alg:edges} to a forest $F\subset Q$.
If we order all edges of $C$ from the vertex $p+\La$ as $e_1,\dots,e_k$, the sum of their translation vectors equals $\vb*{v}$.
We build a path from $p$ to $p+\vb*{v}$ in $G$ by finding a unique edge $[p,p_1]\subset G$ that projects to $e_1$, then a unique edge $[p_1,p_2]\subset G$ that projects to $e_2$ and so on until we cover all $e_1,\dots,e_k$ and arrive at $p+\vb*{v}$. 
\end{proof}

\begin{rmk}
\label{rmk:onus}
The paper \cite{onus2022quantifying} discusses connected components of a periodic graph $K$ in terms of homology, namely Theorem~1(1) proves that
$H_0(K)$ has a basis of $\sum_{i=1}^N [\mathbb{Z}^d : W_{Q_i}]$ elements, see details in their section 3.1, but without describing an algorithm for finding such a basis.
Our results complement their approach by proving the time complexity for checking the connectivity of a dynamic periodic Euclidean graph in Theorem~\ref{thm:bridge_length_time} whilst keeping track of its connected components. 
\end{rmk}

\begin{lem}[ignored edges]
\label{lem:ignored_edges}
Let an edge $e$ 
be a $\La$-equivalence class of a straight-line edge $[p,q]+\La$ in a lifted periodic graph $G$ for some points $p,q\in S$.
If Algorithm~\ref{alg:bridge_length} does not add the edge $e$ to a labelled quotient graph $Q$, then the points $p,q$ are already connected by a path in the graph $G\subset\R^n$ lifted from $Q$ by Lemma~\ref{lem:lifting}. 
\end{lem}
\begin{proof}
The loop stage in Algorithm~\ref{alg:bridge_length} ignores
an edge $e$ in the cases below.
\smallskip

\noindent
Case 1.
The edge $e$ forms a cycle in $Q$ whose cycle sum is the zero vector in $\Z^n$. 
\smallskip

\noindent
Case 2.
The edge $e$ forms a cycle whose cycle sum equals an integer linear combination of pre-existing \emph{cycle sums} from the translational set $B$.
\smallskip

In both cases, we have either one cycle (in case 1) containing $e$, whose cycle sum is $0\in\Z^n$, or several cycles (in case 2), one (up to multiplicity) of which contains $e$, whose total sum of translational vectors is $0\in\Z^n$.
By Definition~\ref{dfn:LQG} each edge of $Q$ involved in this zero sum can be lifted to a straight-line edge in the graph $G\subset\R^n$.
\smallskip

If we start from the given point $p\in S$, a cycle in $Q$ and its sum 0 of translational vectors guarantees that the sequence of the lifted edges in $G$ finishes at the same point $p$ and hence forms a cycle $C$.
This cycle $C$ has the edge $[p,q]$ whose exclusion keeps the points $p,q\in S$ connected by the path in $C$ that is complementary to $[p,q]$.
\end{proof}

\begin{thm}
\label{thm:bridge_length_algorithm}
Algorithm \ref{alg:bridge_length} finds the bridge length $\be(S)$ from Definition~\ref{dfn:bridge_length} for any periodic point set $S\subset\R^n$ with a motif $M$ of points given in a basis $\vb*{v_1},\dots,\vb*{v_n}$.
\end{thm}
\begin{proof}
Within Algorithm \ref{alg:bridge_length}, let $d$ be the length of the last added edge $e$ after which both termination conditions finally hold.
By Lemma~\ref{lem:ignored_edges} all ignored edges do not create extra connections in the graph $G$. 
By Lemmas~\ref{lem:factors1} and~\ref{lem:tree_representatives} the graph $G$ obtained before adding the last edge $e$ is disconnected.
Lemma~\ref{lem:sufficiency} guarantees that, when $e$ is added, the graph $G$ becomes connected. 
Because Algortihm~\ref{alg:edges} yields edges in increasing order, $e$ is the shortest edge that could have this property, so the bridge length is $\beta(S)=d$.
\end{proof}

Theorem~\ref{thm:bridge_length_time} has a rough upper bound assuming that the Smith Normal Form $\SNF(A)$ of an integer $n\times m$ matrix $A$ is re-computed for every iteration in time $O(N)$.
This time was estimated in \cite{giesbrecht1995fast} as $O^\sim(n^{\omega-1}m\cdot M(n \log||A||))$, where $||A||=\max\limits_{i,j}|A_{i,j}|$ and $A_{i,j}$ denotes the element of $A$ in the row $i$ and column $j$.
Here $M(t)$ bounds the cost of multiplying two $t$-bit integers, and $\omega \leq 2.372$ is the exponent for matrix multiplication: two $n\times n$ matrices can be multiplied in time $O(n^{\omega})$, see \cite{williams2024new}.
The “soft-Oh” simplifies the complexity up to logarithmic factors, so $f=O^\sim(G)$ if and only if $f = O(g \log^c g)$ for a constant $c>0$.
\smallskip

To speed up Algorithm~\ref{alg:bridge_length} in practice, the Smith Normal Form can be updated at every iteration instead of recomputing from scratch, see details in appendix~\ref{app:SNF}.

\begin{proof}[\textbf{Proof of Theorem}~\ref{thm:bridge_length_time}] 
Algorithm \ref{alg:bridge_length} solves Problem \ref{pro:bridge_length} by Theorem~\ref{thm:bridge_length_algorithm}.
It remains to show that the time complexity of Algorithm \ref{alg:bridge_length} is $O(m^2 a(U)^nN)$.
Algorithm \ref{alg:bridge_length} has the initialisation of a constant time $O(1)$ and the loop stage.
We will multiply an upper bound for the number of loops by the time complexity of each loop.
\smallskip

One loop in Algorithm~\ref{alg:bridge_length} contains at most the following checks. 
\begin{itemize}
  \item \emph{(Cycle)} Does adding an edge $e$ to a forest $F$ create a cycle? 
  \item \emph{(Combination)} Is the cycle sum an integer combination of previous cycle sums? 
  \item \emph{(Termination)} After appending a cycle sum $c$ to the translational matrix $A$ and 
  calculating $\SNF(A)$, does $A$ have $n$ invariant factors equal to 1?
\end{itemize}

The condition \emph{Cycle} is checked by a depth-first search $O(m)$, see \cite{sedgewick2011algorithms}[4.1].
The condition \emph{Combination} is equivalent to 'Has $\SNF(A)$ changed?', and \emph{Termination} is equivalent to 'Is the product of invariant factors of $A$ equal to 1?'.
So both conditions can be jointly checked in time $O(N)$ needed to compute $\SNF(A)$.
\smallskip

The time complexity of $\SNF(A)$ dominates all other steps in Algorithm \ref{alg:bridge_length}, so we will use $O(N)$ to represent the complexity of a single loop iteration of Algorithm \ref{alg:bridge_length}.
\smallskip
 
Every loop iteration calls Algorithm~\ref{alg:edges}.
If we consider all calls to Algorithm~\ref{alg:edges} as running sequentially, the main loop will run at most $a(U)+1$ times, where $a(U)$ is the aspect ratio from Definition~\ref{dfn:parameters}. 
By Definition~\ref{dfn:parameters}, `$supercell\_size$' must reach at least $a(U)+1$ to ensure that we \emph{yield} all potential edges up to and including $\beta(S)$, i.e., $supercell\_size * h(U) > \beta(S)$.
Each loop runs through the unit cells that are `$supercell\_size$' away from the central cell $U_1$.
By the end, we will have run through and \emph{yielded} at most $(a(U)+1)^n$ unit cells. 
For each unit cell $U_i$, we find all distances between $m$ points in $U_i$ and $m$ points in the central cell.
The required time is $O(m^2)$ between any two cells and hence $O(m^2a(U)^n)$ for all cells.
Algorithm~\ref{alg:bridge_length} does not run for every edge found by Algorithm~\ref{alg:edges}, but we assume this for simplicity. 
The worst-case complexity of this implementation of Algorithm~\ref{alg:bridge_length} is $O(m^2a(U)^nN)$.
\end{proof}


\section{Experiments on real and simulated crystals, and a discussion}
\label{sec:experiments}

\begin{figure}
\includegraphics[width=\textwidth]{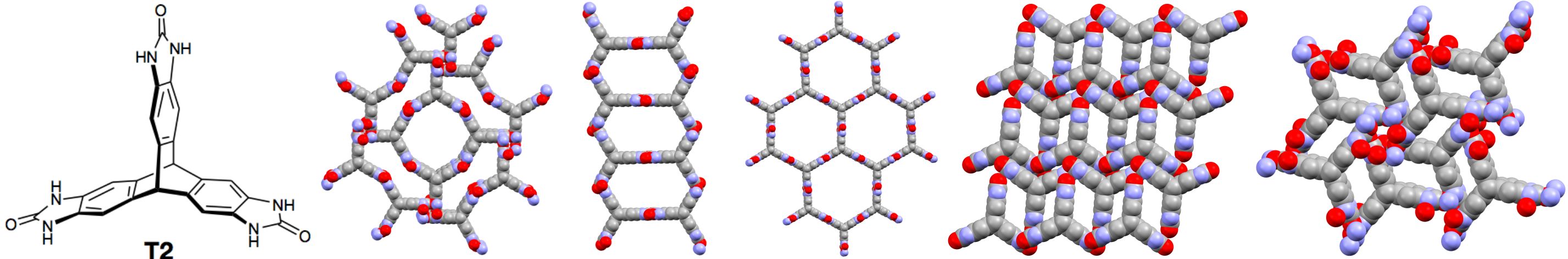}
\caption{
T2 molecule and 5 crystals synthesized from T2.
The first four T2-$\al$, T2-$\be$, T2-$\ga$, T2-$\de$ were reported in \cite{pulido2017functional}, the last T2-$\ep$ in \cite{zhu2022analogy}.}
\label{fig:T2crystals}
\end{figure}

This section discusses experiments computing the exact bridge length $\be(S)$ for 5679 simulated and 5 real nanoporous crystals in Fig.~\ref{fig:T2crystals} reported in Nature paper \cite{pulido2017functional}. 
Table~\ref{tab:T2crystals} contains the bridge lengths computed by Algorithm~\ref{alg:bridge_length} on the crystals from Fig.~\ref{fig:T2crystals} given by their codes in the Cambridge Structural Database (CSD). 
The names of T2 polymorphs refer to the crystalline forms $\al,\be,\ga,\de,\ep$ based on the same molecule T2.
The crystal IDs starting from 6-letter codes in the first column of Table~\ref{tab:T2crystals} refer to the 
Cambridge Structural Database \cite{taylor2019million}.

Note that the polymorph T2-$\ga$ contains four slightly different versions in the CSD (DEBXIT01$\dots$04) because their crystal structures were determined at different temperatures. 
The seven versions DEBXIT01$\dots$07 with the same 6-letter code may look similar even for experts.
The polymorph T2-$\de$ (SEMDIA) was deposited later than others because even the original authors confused this polymorph with earlier crystals.
This confusion was detected by density functions from \cite{edelsbrunner2021density}, computed by \cite{smith2022practical}.
The underlying density invariants turned out to be incomplete by Example~11 in \cite{anosova2022density}, but were explicitly described for all periodic sequences of intervals in $\R$ \cite{anosova2023density}. 
\smallskip

Table~\ref{tab:T2crystals} includes the upper bounds $\be(S)\leq \min\{r(U),2R(S)\}$ from Lemma~3.6(a) in \cite{anosova2025recognition}, see $r(U)$ and $R(S)$ in Definition~\ref{dfn:parameters}.
The run times in Table~\ref{tab:T2crystals} were recorded on a laptop with Intel i5, one 1GHz core, and 8Gb RAM.
\smallskip

The final row contains the averages for 5,679 simulated T2 crystals, which are publicly available in the supplementary materials of \cite{pulido2017functional} and were used for predicting the 5 experimental polymorphs represented by 9 entries in the CSD.
For all crystals in Table~\ref{tab:T2crystals}, the translational matrix size never exceeded 3 columns.
\medskip

\begin{table}
\caption{The exact bridge length $\be(S)$ computed by Algorithm~\ref{alg:bridge_length} and its upper bounds for 
the 9 experimental and 5679 simulated T2 crystals reported by \cite{pulido2017functional}.}
\label{tab:T2crystals}
\begin{tabular}{l|crrrrr}      
 CSD ref codes of &
        number &
        bridge  & 
        upper & 
        upper &
        best upper &
        running  \\ 
        experimental and &
        of atoms  &
        length  & 
        bound  & 
        bound &
        bound over  &
        time,  \\ 
simulated crystals &
in a cell &
$\beta(S)$, \r{A}  & 
 $r(U)$, \r{A}  & 
 $2R(S)$, \r{A} &
 exact $\beta(S)$ &
seconds \\
\hline
 T2-$\al$ NAVXUG & 184 & 2.028 & 22.325 & 15.609 & 7.695 & 4.337\\
        T2-$\be$ DEBXIT05 & 92 & 3.163 & 20.665 & 12.906 & 4.080 & 0.664\\
        T2-$\be$ DEBXIT06 & 92 & 3.188 & 20.694 & 12.884 & 4.042 & 0.657\\
         T2-$\ga$ DEBXIT01 & 92 & 1.879 & 23.224 & 23.366 & 12.358 & 0.706\\
        T2-$\ga$ DEBXIT02 & 92 & 1.926 & 23.226 & 23.375 & 12.061 & 0.636\\
        T2-$\ga$ DEBXIT03 & 92 & 1.902 & 23.230 & 23.373 & 12.216 & 0.653\\
        T2-$\ga$ DEBXIT04 & 92 & 1.970 & 23.290 & 23.448 & 11.824 & 0.649 \\
        T2-$\de$ SEMDIA & 92 & 2.713 & 14.401 & 8.350 & 3.077 & 0.671\\ 
         T2-$\ep$ DEBXIT07 & 92 & 2.062 & 12.608 & 5.707 & 2.768 & 0.641\\
        \hline
        average for all 5679 & 295.8 & 2.293 & 23.306 & 9.110 & 4.064 &  31.653\\
        simulated T2 crystals & & & & & &
\end{tabular}
\end{table}

The real T2 crystals in the CSD have smaller motifs consisting of only 2 or 4 T2 molecules, while simulated T2 crystals contain up to 32 molecules, which makes the running times slower in comparison with real ones, see the last column in Table~\ref{tab:T2crystals}.
\smallskip

More importantly, the exact bridge length $\be(S)$ is 4 times smaller (on average) than its upper bound $\min\{r(U),2R(S)\}$.
The bridge length $\be(S)$ provides the upper bound $\be(S)+2R(S)>\al(S)$ in Lemma~3.6(b) from \cite{anosova2025recognition} for a stable radius $\al$ of atomic clouds that suffices for a complete and continuous isoset invariant of $S$.
\smallskip

This isoset uniquely identifies any periodic crystal $S$ under rigid motion and has a continuous distance metric that has detected thousands of near-duplicate crystals.
Decreasing the upper bound of $\al(S)$ from $4R(S)$ to the smaller value $\be(S)+2R(S)$ by a factor of about 2 decreases the size $m$ of atomic clouds by a factor of $2^3=8$ in $\R^3$.
This size reduction speeds up by several orders of magnitude the algorithms for isosets and their distance metric, which have complexity $O(m^3\log m)$ and $O(m^6)$ in $\R^3$, respectively; see the conclusions of section~5 in \cite{anosova2025recognition}.
\smallskip

The next open problem is an exact computation of the minimal stable radius $\al(S)$, The closely related problem is to compute the \emph{regularity radius} $\rho$ that is the minimum radius with the property that any Delone set with mutually equivalent clusters of the radius $\rho$ is \emph{regular} (periodic with a 1-point  asymmetric), see \cite{baburin2018origin}.
\smallskip

In conclusion, this paper contributes an exact algorithm for the bridge length, a key ingredient for solving the geo-mapping problem for periodic point sets in Geometric Data Science \cite{anosova2021introduction,kurlin2025complete,anosova2026geometric}. 
This problem has been solved for 2D lattices in \cite{kurlin2024mathematics,bright2023geographic,bright2023continuous}, while the 3D case is being finalised in \cite{kurlin2022complete,bright2021complete}. 

\ack{Acknowledgements}.
We thank Jean-Guillaume Eon and Gregory McColm 
for their helpful advice on the early draft of the paper.
This work was supported by the Royal Society APEX fellowship 
APX/R1/231152 and EPSRC New Horizons EP/X018474/1.

\bibliographystyle{iucr}
\bibliography{iucr}

@book{hatcher2011AT,
  address = {Cambridge},
  author = {Hatcher, Allen},
  isbn = {0-521-79160-X; 0-521-79540-0},
  mrclass = {55-01 (55-00)},
  mrnumber = {1867354 (2002k:55001)},
  mrreviewer = {Donald W. Kahn},
  pages = {xii+544},
  publisher = {Cambridge University Press},
  title = {Algebraic topology},
  username = {mwpb479},
  year = 2002
}

@inproceedings{delone1976local,
  title={A local criterion for regularity of a system of points},
  author={Delone, Boris Nikolaevich and Dolbilin, Nikolai Petrovich and Shtogrin, Mikhail Ivanovich and Galiulin, Ravil V},
  booktitle={Dokl. Akad. Nauk SSSR},
  volume={227},
  number={1},
  pages={19--21},
  year={1976}
}

@article{baburin2018origin,
  title={On the origin of crystallinity: a lower bound for the regularity radius of Delone sets},
  author={Baburin, Igor A and Bouniaev, Mikhail and Dolbilin, Nikolay and Erokhovets, Nikolay Yu and Garber, Alexey and Krivovichev, Sergey V and Schulte, Egon},
  journal={Acta Crystallographica Section A},
  volume={74},
  number={6},
  pages={616--629},
  year={2018}
}

@misc{brock2021change,
author={Carolyn P. Brock},
year={2021},
howpublished = {https://www.iucr.org/news/newsletter/etc/articles?issue=151351\&result\_138339\_result\_page=17},
title = {Change to the definition of "crystal"}
}

@article{anosova2024importance,
	author = {Olga Anosova and Vitaliy Kurlin and Marjorie Senechal},
	title = {The importance of definitions in crystallography},
	journal = {IUCrJ},
	volume = {11},
	issue = {4},
	pages = {453-463},
	doi = {10.1107/S2052252524004056},
	year = {2024}
}

@inproceedings{williams2024new,
  title={New bounds for matrix multiplication: from alpha to omega},
  author={Williams, Virginia and others},
  booktitle={Symposium on Discrete Algorithms},
  pages={3792--3835},
  year={2024},
  organization={SIAM}
}

@article{baladi2005euclidean,
  title={Euclidean algorithms are Gaussian},
  author={Baladi, Viviane and Vall{\'e}e, Brigitte},
  journal={Journal of Number Theory},
  volume={110},
  number={2},
  pages={331--386},
  year={2005},
  publisher={Elsevier}
}

@inproceedings{giesbrecht1995fast,
  title={Fast computation of the Smith normal form of an integer matrix},
  author={Giesbrecht, Mark},
  booktitle={Symposium on symbolic and algebraic computation},
  pages={110--118},
  year={1995}
}

@book{sedgewick2011algorithms,
  title={Algorithms},
  author={Sedgewick, Robert and Wayne, Kevin},
  year={2011},
  publisher={Addison-wesley professional}
}

@article{edelsbrunner2024merge,
  title={Merge Trees of Periodic Filtrations},
  author={Edelsbrunner, Herbert and Heiss, Teresa},
  journal={arXiv:2408.16575},
  year={2024}
}

@article{widdowson2024continuous,
	author = {Daniel Widdowson and Vitaliy Kurlin},
	title = {Continuous invariant-based maps of the Cambridge Structural Database},
	journal = {Crystal Growth and Design},
	volume = {24},
	issue = {13},
	pages = {5627–5636},
	year = {2024},
	doi = {10.1021/acs.cgd.4c00410}
}

@inproceedings{smith2022practical,
  title={A practical algorithm for degree-k Voronoi domains of three-dimensional periodic point sets},
  author={Phil Smith and Vitaliy Kurlin},
  booktitle={Lecture Notes in Computer Science},
  volume={13599},
  pages={377-391},
  year={2022}
}

@article{anosova2021introduction,
  title={Introduction to Periodic Geometry and Topology},
  author={O Anosova and V Kurlin},
  journal={arXiv:2103.02749},
  year={2021}
}

@article{anosova2026geometric,
  title={Geometric Data Science},
  author={O Anosova and V Kurlin},
  journal={arXiv:2512.05040},
  year={2025}
}

@inproceedings{anosova2021isometry,
  author={O Anosova and V Kurlin},
  title={An isometry classification of periodic point sets},
  booktitle={Lecture Notes in Computer Science (Proceedings of Discrete Geometry and Mathematical Morphology)},
  volume = {12708},
  pages={229-241},
  year = {2021}
  }

@article{anosova2025recognition,
 	 title={Recognition of near-duplicate periodic patterns by continuous metrics with approximation guarantees},
  	author={Olga Anosova and Daniel Widdowson and Vitaliy Kurlin},
  	journal={Pattern Recognition},
  	volume={171},
	number={112108},
	year={2025}
}

@article{dolbilin2018delone,
  title={Delone Sets in $\R^3$ with $2R$-Regularity Conditions},
  author={Dolbilin, NP},
  journal={Proceedings of the Steklov Institute of Mathematics},
  volume={302},
  number={1},
  pages={161--185},
  year={2018},
  publisher={Springer}
}

@inproceedings{dolbilin2015delone,
  title={Delone sets: local identity and global symmetry},
  author={Dolbilin, Nikolay},
  booktitle={Geometry and symmetry conference},
  pages={109--125},
  year={2015},
  organization={Springer}
}

@inproceedings{dolbilin1976local,
  title={Local properties of discrete regular systems},
  author={Dolbilin, Nikolay Petrovich},
  booktitle={Doklady Akademii Nauk},
  volume={230},
  number={3},
  pages={516--519},
  year={1976}
}

@book{newman1972integral,
  title={Integral matrices},
  author={Newman, Morris},
  year={1972},
  publisher={Academic Press}
}

@book{cohn1985free,
  title={Free rings and their relations},
  author={Cohn, Paul Moritz},
  publisher={Academic Press},
  year={1985}
}

@article{pulido2017functional,
  title={Functional materials discovery using energy--structure maps},
  author={Pulido, Angeles and others},
  journal={Nature},
  volume={543},
  pages={657-664},
  year={2017}
}

@article{zhu2022analogy,
  title={Analogy Powered by Prediction and Structural Invariants},
  author={Q Zhu and others},
  journal={J Amer. Chem. Soc.},
  volume={144},
  pages={9893–9901},
  year={2022}
}

@article{eon2016topological,
  title={Topological features in crystal structures: a quotient graph assisted analysis of underlying nets and their embeddings},
  author={Eon, J-G},
  journal={Acta Crystallographica Section A},
  volume={72},
  number={3},
  pages={268--293},
  year={2016},
  publisher={International Union of Crystallography}
}

@article{eon2016vertex,
  title={Vertex-connectivity in periodic graphs and underlying nets of crystal structures},
  author={Eon, J-G},
  journal={Acta Crystallographica Section A},
  volume={72},
  number={3},
  pages={376--384},
  year={2016},
  publisher={International Union of Crystallography}
}

@article{eon2011euclidean,
  title={Euclidean embeddings of periodic nets: definition of a topologically induced complete set of geometric descriptors for crystal structures},
  author={Eon, J-G},
  journal={Acta Crystallographica Section A},
  volume={67},
  number={1},
  pages={68--86},
  year={2011}
}

@article{mccolm2024realizations,
  title={Realizations of crystal nets. I (Generalized) derived graphs},
  author={McColm, Gregory},
  journal={Acta Crystallographica Section A},
  volume={80},
  number={1},
  year={2024}
}

@article{taylor2019million,
  title={A million crystal structures: The whole is greater than the sum of its parts},
  author={Taylor, Robin and Wood, Peter A},
  journal={Chemical reviews},
  volume={119},
  number={16},
  pages={9427--9477},
  year={2019},
  publisher={ACS Publications}
}

@inproceedings{edelsbrunner2021density,
  title={The Density Fingerprint of a Periodic Point Set},
  author={Edelsbrunner, H and others},
  booktitle={Proceedings of SoCG},
  pages={32:1--32:16},
  year={2021}
  }

@book{van2003algebra,
  title={Algebra},
  author={Van der Waerden, Bartel Leendert},
  volume={1},
  year={2003},
  publisher={Springer Science \& Business Media}
}

@article{widdowson2022average,
  title={Average Minimum Distances of periodic point sets - foundational invariants for mapping all periodic crystals},
  author={Widdowson, Daniel and others},
  journal={MATCH Commun. Math. Comput. Chem.},
  volume={87},
  pages={529-559},
  year={2022}
}

@article{widdowson2026pointwise,
	author = {Daniel Widdowson and Vitaliy Kurlin},
	title = {Pointwise Distance Distributions for detecting near-duplicates in large materials databases},
	journal = {SIAM J Applied Mathematics, 10.1137/25M1736657},
	year = {2026}
}

@article{widdowson2022resolving,
  title={Resolving the data ambiguity for periodic crystals}, 
  author={Daniel Widdowson and Vitaliy Kurlin},
  journal={NeurIPS},
  pages={24625-24638},
  volume={35},
  year={2022}
}

@article{kurlin2024mathematics,
  title={Mathematics of 2{D} lattices}, 
  author={Vitaliy Kurlin},
  journal={Foundations of Computational Mathematics},
  volume={24}, 
  pages={805–863},
  year={2024}
}

@article{bright2021complete,
  title={A complete and continuous map of the lattice isometry space for all 3{D} lattices},
  author={Bright, Matthew and Cooper, Andrew and Kurlin, Vitaliy},
  journal={arXiv:2109.11538},
  year={2021}
}

@article{bright2023geographic,
  title={Geographic-style maps for 2{D} lattices}, 
  author={Matthew Bright and Andrew Cooper and Vitaliy Kurlin},
  journal={Acta Cryst A},
  volume ={79},
  pages={1-13},
  year={2023}
}

@article{bright2023continuous,
  title={Continuous chiral distances for 2-dimensional lattices}, 
  author={Matthew Bright and Andrew Cooper and Vitaliy Kurlin},
  journal={Chirality},
  volume={35},
  issue={12},
  pages={920-936},
  year={2023}
}

@article{bouniaev2017regular,
  title={Regular and Multi-regular t-bonded Systems},
  author={Bouniaev, M. and Dolbilin, N.},
  journal={J. Information Processing},
  volume={25},
  pages={735--740},
  year={2017}
}

@article{dolbilin2019regular,
  title={Regular t-bonded systems in {R}$^3$},
  author={Dolbilin, N and Bouniaev, M},
  journal={European Journal of Combinatorics},
  volume={80},
  pages={89--101},
  year={2019}
}

@article{dolbilin1998multiregular,
  title={Multiregular point systems},
  author={Dolbilin, N. and Lagarias, J and Senechal, M.},
  journal={Discrete Comp. Geometry},
  volume={20},
  number={4},
  pages={477--498},
  year={1998}
}

@inproceedings{delone1973three,
  title={On three successive minima of a three-dimensional lattice},
  author={Delone, Boris Nikolaevich and Galiulin, Ravil V and Dolbilin, Nikolay Petrovich and Zalgaller, Viktor Abramovich and Shtogrin, Mikhail Ivanovich},
  booktitle={Doklady Akademii Nauk},
  volume={209},
  number={1},
  pages={25--28},
  year={1973}
}

@article{kurlin2022complete,
 title={A complete isometry classification of 3-dimensional lattices},
 author={V Kurlin},
  journal={arxiv:2201.10543},
 year={2022}
}

@article{kurlin2025complete,
	author = {Vitaliy Kurlin},
	title = {Complete and continuous invariants of 1-periodic sequences in polynomial time},
	journal = {SIAM Journal on Mathematics of Data Science},
	volume = {7},
	issue = {4},
	pages = {1643-1663},
	year = {2025}
}

@inproceedings{anosova2022density,
  title={Density functions of periodic sequences},
  author={Olga Anosova and Vitaliy Kurlin},
  booktitle={Lecture Notes in Computer Science (Proceedings of Discrete Geometry and Mathematical Morphology)},
  volume={13493},
  pages={395-408},
  year={2022}
}

@article{anosova2023density,
  title={Density functions of periodic sequences of continuous events},
  author={Anosova, Olga and Kurlin, Vitaliy},
  journal={Journal of Mathematical Imaging and Vision},
  volume={65},
  pages={689–701},
  year={2023}
}

@article{onus2022quantifying,
  title={Quantifying the homology of periodic cell complexes},
  author={Onus, Adam and Robins, Vanessa},
  journal={arXiv:2208.09223},
  year={2022}
}

@article{chung1984nomenclature,
  title={Nomenclature and generation of three-periodic nets: the vector method},
  author={Chung, Sui Jin and Hahn, Th and Klee, WE},
  journal={Acta Crystallographica Section A},
  volume={40},
  number={1},
  pages={42--50},
  year={1984},
  publisher={International Union of Crystallography}
}

@inproceedings{cohen1990recognizing,
  title={Recognizing properties of periodic graphs},
  author={Cohen, Edith and Megiddo, Nimrod},
  booktitle={Applied geometry and discrete mathematics},
  pages={135--146},
  year={1990}
}

\vspace*{-5mm}

\appendix{}
\section{A faster 'online' algorithm for the Smith Normal Form}
\label{app:SNF}
  
 A different way of checking the \emph{Termination} condition is to append columns to $A$ in an 'online' fashion.
This avoids the need to calculate the Smith Normal Form from scratch every time (or often at all), and reduces the complexity to a time close to $O(m^{\omega}\cdot E\cdot n)$, where $\omega \leq 2.372$ is the exponent for matrix multiplication \cite{williams2024new}, and 
$O(E)$ is the complexity of the Extended Euclidean Algorithm \cite{baladi2005euclidean}. 
As this reduction in complexity is dominated by the price of populating the edges with Algorithm \ref{alg:edges}, this will be irrelevant for most use cases (and is not used in the experiments shown later).
As a use case involves, say, a larger or higher-dimensional pre-populated set of edges, this algorithm becomes more necessary.

Recalling from Definition~\ref{dfn:SNF} that the diagonal of the Smith Normal Form $\SNF(A)=LAR$ is made up of the invariant factors of $A$.
To progressively calculate $\SNF(A)$, we must only keep track of the right-multiplying unimodular matrix $R$, and the invariant factors themselves, which form a vector $\vb*{f}=(f_1,\dots,f_n) \in \mathbb{Z}^n$.
To run the main algorithm here, we do have to begin with a matrix with n integer linearly independent rows. 
'Adding' a vector $\vb*{v}$ to $\vb*{f}$ is where the process changes. We treat $R$ and $\vb*{f}$ as mutable, meaning each value is not necessarily fixed to its original assignment.
The first step is to define $x := \vb*{v}*R$, then we find $g_i=gcd(x_i,f_i)$. If $f_i = g_i$ (i.e., $f_i\ divides\ x_i$), we can continue with $i:=i+1$, with no need to change $R$ as it only keeps track of columns (for context, if we were keeping track of $L$, too, we would have to subtract the $i$-{th} row from the last row $x_i/f_i$ times).

If $f_i$ divides $x_i$ for all $i$, we would know that including the vector changes nothing, therefore the relative edge is also irrelevant and can be discarded (this reduces the complexity of most of the \emph{Termination} condition from $O(N)$ to $O(n^{\omega} + log^2(n)$).

However, if $g_i<f_i$, then $f_i$ not only becomes $g_i$, but we also know that $\SNF(A)$ will change and that we must add the edge relative to $\vb*{v}$. 
We must also alter $R$, accounting for the fact that $F$ represents the diagonal of a matrix. 
We can do this by any typical process of 'changing the pivot' in the $\SNF$ algorithm, ensuring that we update $R$ in tandem.
As accounting for the previous values of $i$ is trivial, it is the worst-case equivalent to calculating the $\SNF$ of an $(n-i)\times(n-i)$ matrix in time $O(N_{n-i})$, which improves upon the naive calculation of $\SNF$ from scratch upon every alteration of $A$.

\begin{lem}
Updating the Smith Normal Form as above preserves its properties.
\end{lem}
\begin{proof}
As we only alter with elementary row and column operations, this preserves the Smith Normal Form. By multiplying the to-be-added row $\vb*{v}$ by $R$ before concatenating it as a new row to $F$, it is the same as performing those same elementary column operations upon a new matrix: $[\vb*{A_0},..,\vb*{A_n},\vb*{v}]$ (i.e. $\vb*{v}$ concatenated as a row onto $A$). 

We then continue to perform only elementary row and column operations, and we end with a matrix that satisfies the conditions of an $\SNF$ noted in Definition \ref{dfn:SNF}.
\end{proof}   

To discuss this process any further is beyond the scope of this paper, though there are still some small tricks that take advantage of the way the 'new' rows for consideration are intrinsically related to $\vb*{v}$, and how $f_{i+1}$ divides $f_i$.






\end{document}